\documentclass[12pt,a4paper]{article}

\usepackage{latexsym}
\usepackage{amsmath}
\usepackage{amsfonts}
\usepackage{amssymb}

\begin{document}

\vspace*{0.5cm}
\begin{center}
{\Large \bf On the self-adjointness of certain reduced
Laplace--Beltrami operators}
\end{center}

\vspace{0.2cm}

\begin{center}
L. FEH\'ER${}^{a}$ and B.G. PUSZTAI${}^b$ \\

\bigskip

${}^a$Department of Theoretical Physics, MTA  KFKI RMKI\\
1525 Budapest 114, P.O.B. 49,  Hungary, and\\
Department of Theoretical Physics, University of Szeged\\
Tisza Lajos krt 84-86, H-6720 Szeged, Hungary\\
e-mail: lfeher@rmki.kfki.hu

\bigskip
${}^b$Centre de recherches math\'ematiques, Universit\'e de Montr\'eal\\
C.P. 6128, succ. centre ville, Montr\'eal, Qu\'ebec, Canada H3C 3J7, and\\
Department of Mathematics and Statistics, Concordia University\\
1455 de Maisonneuve Blvd. West, Montr\'eal,  Canada H3G 1M8\\
e-mail: pusztai@CRM.UMontreal.CA

\end{center}

\vspace{0.2cm}

\begin{abstract}
The self-adjointness of the reduced Hamiltonian operators arising
from the Laplace--Beltrami operator of a complete Riemannian
manifold through quantum Hamiltonian reduction based on  a compact
isometry group is studied. A simple sufficient condition is provided
that guarantees the inheritance of essential self-adjointness onto a
certain class of restricted operators and allows us to conclude the
self-adjointness of the reduced Laplace--Beltrami operators in a
concise way. As a consequence, the self-adjointness of spin
Calogero--Sutherland type reductions of `free' Hamiltonians under
polar actions of compact Lie groups follows immediately.
\end{abstract}

\bigskip


\newpage

\section{Introduction}

The goal of this work is to verify the self-adjointness of certain reduced
Hamiltonians obtained from quantum Hamiltonian reduction.
The importance of the reduction method
(see e.g.~\cite{Lands, TanimuraIwai})
mainly stems from the fact that under suitable symmetries
the Hilbert space of a quantum system  can be decomposed into
invariant subspaces simplifying the diagonalization of the
Hamiltonian. Another attractive perspective
comes from the theory of integrable systems, since
many  integrable models arise as Hamiltonian reductions of canonical
 `free' systems that are solvable due to their symmetries.
The self-adjointness of the reduced Hamiltonians is in general
necessary for the usefulness of the method.

If a compact Lie group $G$ acts smoothly on a complete Riemannian manifold
$(Y, \eta)$ by isometries, then  $G$ becomes a symmetry group of
the quantum system of the
free particle on $Y$.   We take the free Hamiltonian to be
just the Laplace--Beltrami
operator, $\Delta_Y$, whose self-adjointness on a natural domain
is guaranteed  by
the completeness of the underlying metric $\eta$.
The system breaks up into
$G$-invariant subspaces labeled by unitary irreducible
representations $(\rho, V)$ of $G$, and
the quantum Hamiltonian reduction gives rise to
reduced Hamiltonians, $\Delta_\rho$, associated with
these invariant subspaces.

In a recent paper \cite{FP_polar}, we studied quantum Hamiltonian reductions
of Laplace--Beltrami operators on complete Riemannian manifolds under
isometric actions of compact Lie groups that permit
the introduction of generalized polar coordinates as defined in \cite{PT}.
In this case we derived an explicit formula for the reduced
Laplace--Beltrami operators and stated their essential
self-adjointness
on certain domains without detailed proof.
Here, we provide the missing proof.
Many examples of these reduced Hamiltonians yield
integrable systems of spin Calogero--Sutherland type
\cite{OPII,EFK,Obl,FP1,FP3}, and
the present paper is part of our program aimed at exploring these
interesting integrable systems in detail.

The content of the rest of the paper is as follows.
In Section 2 we enquire whether the property of essential self-adjointness
descends onto restrictions of essentially self-adjoint operators,
and provide a sufficient condition under which essential
self-adjointness of certain restrictions can be maintained.
By using this condition, in Section 3 we prove
the essential
self-adjointness of the reduced Laplace--Beltrami operators
in the general case of
smooth isometric actions of compact Lie groups.
We then apply this result to the important special case when $G$
acts on $Y$ in a \emph{polar} manner in the
sense of Palais and Terng \cite{PT}.
In Section 4 we briefly recall from \cite{FP_polar}
 the explicit formula of the reduced Laplace--Beltrami
 operators in the polar case, and confirm their essential self-adjointness.
 Finally, we give our conclusions in Section 5.

\section{Inheritance of essential self-adjointness upon restriction}

Let $A \colon {\mathcal{D}}(A) \rightarrow {\mathcal{H}}$ be a densely defined
symmetric linear operator on a Hilbert space
$({\mathcal{H}}, \langle \, , \rangle)$. That is, the domain of $A$,
${\mathcal{D}}(A)$, is a dense linear subspace of ${\mathcal{H}}$, and
$A$ satisfies the relation
\begin{equation}
\langle A f_1, f_2 \rangle = \langle f_1, A f_2 \rangle
\quad
\forall f_1, f_2 \in {\mathcal{D}}(A).
\label{2.1}\end{equation}
Let $S \subset {\mathcal{H}}$ be an \emph{invariant
linear subspace} of $A$, which means that
$S \subset {\mathcal{D}}(A)$ and $A S \subset S$.
These two assumptions guarantee that the restricted operator
\begin{equation}
B := A|_S \colon S \rightarrow S
\label{2.2}\end{equation}
is well defined.
We can regard  $B$ as a densely defined symmetric operator
on the Hilbert space ${\mathcal{M}}:= \bar{S}$, where $\bar{S}$ denotes
the closure of $S$ in ${\mathcal{H}}$.

Essentially self-adjoint operators
admit unique self-adjoint extensions by their closures.
It is easy to see that essential self-adjointness of $A$ does not imply,
in general, the essential self-adjointness of the restricted operator
$B = A|_S$.  Below we present a sufficient condition under which
the property of essential self-adjointness descends onto the
restricted operator.

Recall that the domain of the adjoint of $A$ is defined by the
subspace
\begin{eqnarray}
&&{\mathcal{D}}(A^*) = \{g \, | \, g \in {\mathcal{H}} \mbox{ for which }
\exists k \in {\mathcal{H}} \mbox{ such that } \nonumber\\
&&\qquad\qquad\qquad
\langle g, A f \rangle = \langle k, f\rangle \mbox{ for every } f
\in {\mathcal{D}}(A)\},
\label{2.3}\end{eqnarray}
and the adjoint of $A$ is the linear operator
\begin{equation}
A^* \colon {\mathcal{D}}(A^*) \rightarrow {\mathcal{H}},
\quad
g \mapsto A^* g := k,
\label{2.4}\end{equation}
where $k \in {\mathcal{H}}$ is the vector appearing in the definition of
${\mathcal{D}}(A^*)$.
The `deficiency subspaces',
$L_A^\pm \subset {\mathcal{D}}(A^*)$, of
the operator $A$ are the sets
\begin{equation}
L_A^\pm := \mathrm{Ker}(A^* \mp {\mathrm{i}}) =
 \{ g \, | \, g \in {\mathcal{D}}(A^*), A^* g = \pm {\mathrm{i}} g \}.
\label{2.5}\end{equation}
As is well known (see e.g.~\cite{RS_II}),
essential self-adjointness of a densely defined
symmetric linear operator $A$ can be characterized by its
deficiency subspaces $L^\pm_A$.
Namely, $A$ is essentially self-adjoint if and only if its deficiency
subspaces are trivial,  $L_A^\pm = \{ 0 \}$.
We can consider the adjoint
$B^* \colon {\mathcal{D}}(B^*) \rightarrow {\mathcal{M}}$ of
the restricted operator $B$ (\ref{2.2}), and its deficiency subspaces, too.
Next we relate the adjoint of $B$ to the adjoint of $A$.
\\[2mm]

\noindent
\textbf{Lemma 2.1.}
\emph{
Suppose that the domain of $A$ and the $A$-invariant linear subspace $S$
satisfy the following additional compatibility condition
\begin{equation}
P_{\mathcal{M}} {\mathcal{D}}(A) \subset S,
\label{2.6}\end{equation}
where $P_{\mathcal{M}} \colon {\mathcal{H}} \rightarrow {\mathcal{M}}$ denotes
the orthogonal projection onto the closed subspace ${\mathcal{M}} = \bar{S}$.
Then $A^*$ is an \emph{extension} of $B^*$, $B^* \subset A^*$, that is,
${\mathcal{D}}(B^*) \subset {\mathcal{D}}(A^*)$ and
$A^*|_{{\mathcal{D}}(B^*)} = B^*$. }\\[2mm]

\noindent
\textbf{Proof.}
First, take an arbitrary vector $f \in {\mathcal{D}}(A)$, and write it
 according to the orthogonal decomposition
${\mathcal{H}} = {\mathcal{M}} \oplus {\mathcal{M}}^\perp$ as
$f = P_{\mathcal{M}} f + P_{{\mathcal{M}}^\perp} f$.
Using the condition (\ref{2.6}) we see that
$P_{\mathcal{M}} f \in S \subset {\mathcal{D}}(A)$,
therefore
$P_{{\mathcal{M}}^\perp} f = (f - P_{\mathcal{M}} f) \in {\mathcal{D}}(A)$.
Moreover, for every $u \in S$ we have
$\langle u, A(P_{{\mathcal{M}}^\perp} f)\rangle = \langle A u, P_{{\mathcal{M}}^\perp} f \rangle = 0$,
whence we obtain
$A(P_{{\mathcal{M}}^\perp} f) \in S^\perp = {\mathcal{M}}^\perp$.

Second, let $g \in {\mathcal{D}}(B^*)$ be an arbitrary vector. Then for every
$f \in {\mathcal{D}}(A)$ we obtain
\begin{eqnarray}
&&\langle g, A f \rangle
= \langle g, A P_{{\mathcal{M}}} f + A P_{{\mathcal{M}}^\perp} f \rangle
= \langle g, A(P_{\mathcal{M}} f) \rangle
= \langle g, B(P_{\mathcal{M}} f)\rangle
\nonumber\\
&&\quad
= \langle B^* g, P_{\mathcal{M}} f \rangle
= \langle B^*g, P_{\mathcal{M}} f \rangle +
\langle B^* g, P_{{\mathcal{M}}^\perp} f \rangle
= \langle B^* g, f \rangle.
\label{2.7}\end{eqnarray}
Therefore $g \in {\mathcal{D}}(A^*)$ and $A^* g = B^* g$.
\rule{5pt}{5pt}
\\[2mm]

\noindent
\textbf{Corollary 2.1.}
\emph{
Under the above assumptions on $S$ and ${\mathcal{D}}(A)$, for the deficiency
subspaces of the operators $A$ and $B = A|_S$ we have the inclusion
relations $L_B^\pm \subset L_A^\pm$,
therefore if $A$ is essentially self-adjoint, so is its restriction $B$.
}

\section{Reduction of the Laplace--Beltrami operator}

Let $Y$ be a smooth, connected, complete Riemannian
manifold with metric $\eta$.
The restriction of the corresponding Laplace--Beltrami operator, $\Delta_Y$,
onto the space of
the complex valued compactly supported smooth functions,
\begin{equation}
\Delta_Y^0 := \Delta_Y|_{C_c^\infty(Y)}
\colon C_c^\infty(Y) \rightarrow C_c^\infty(Y),
\label{3.1}\end{equation}
is an essentially self-adjoint linear operator
of the Hilbert space $L^2(Y, {\mathrm{d}} \mu_Y)$, where $\mu_Y$ denotes the
  measure generated by the Riemannian volume form
(see e.g. \cite{Lands} and references therein). Suppose that a
\emph{compact} Lie group $G$ acts on $(Y, \eta)$ by isometries.
More precisely,
we are given a smooth left-action
\begin{equation}
\phi \colon G \times Y \rightarrow Y,
\quad
(g, y) \mapsto \phi(g, y) = \phi_g(y) = g . y
\label{3.2}\end{equation}
of $G$, such that
$\phi_g^* \eta = \eta$ for every $g \in G$.
The measure $\mu_Y$ inherits the  $G$-invariance. Now
take a finite dimensional continuous unitary irreducible
representation
$(\rho, V)$ of $G$, where $V$ is a complex vector space with inner product
$(\, , )_V$. By simply acting componentwise, the operator $\Delta_Y^0$
extends onto the complex vector space of the $V$-valued compactly
supported smooth functions, $C_c^\infty(Y, V)$.
This gives the essentially self-adjoint operator
\begin{equation}
\Delta_Y^0 \colon C_c^\infty(Y, V) \rightarrow C_c^\infty(Y, V)
\label{3.3}\end{equation}
of the Hilbert space $L^2(Y, V, {\mathrm{d}} \mu_Y)$.
Because of the $G$-symmetry of the metric  $\eta$,
the set
\begin{equation}
C_c^\infty(Y, V)^G := \{ F \, | \, F \in C_c^\infty(Y, V),
F \circ \phi_g = \rho(g) \circ F \quad \forall g \in G \}
\label{3.4}\end{equation}
of the $V$-valued, compactly supported \emph{$G$-equivariant} smooth
functions is an invariant linear subspace of
$\Delta_Y^0$, i.e.,
$C_c^\infty(Y, V)^G \subset C_c^\infty(Y, V)$ and
$\Delta_Y^0 C_c^\infty(Y, V)^G \subset C_c^\infty(Y, V)^G$.
\\[2mm]

\noindent
\textbf{Proposition 3.1.}
\emph{
The restriction of
$\Delta_Y^0$ (\ref{3.3})
onto $C_c^\infty(Y, V)^G$,
\begin{equation}
\Delta_\rho := \Delta_Y^0|_{C_c^\infty(Y, V)^G}
\colon C_c^\infty(Y, V)^G \rightarrow C_c^\infty(Y, V)^G,
\label{3.5}\end{equation}
is a densely defined, symmetric, essentially self-adjoint linear operator on
the Hilbert space
$L^2(Y, V, {\mathrm{d}} \mu_Y)^G$
of the  square-integrable $G$-equivariant functions.
}\\[2mm]

\noindent
\textbf{Proof.}
Notice that the closure of
$C_c^\infty(Y, V)^G$
in $L^2(Y, V, {\mathrm{d}} \mu_Y)$ equals
$L^2(Y, V, {\mathrm{d}} \mu_Y)^G$.
Then it is enough to verify the compatibility
condition (\ref{2.6}), which  in our case requires proving that
$P_{\mathcal{M}} C_c^\infty(Y, V) \subset C_c^\infty(Y, V)^G$
with
${\mathcal{M}} := L^2(Y, V, {\mathrm{d}} \mu_Y)^G$.

For each $F \in C_c^\infty(Y, V)$,  let us define the function
$\hat F: Y\to V$
by averaging over $G$,
\begin{equation}
Y \ni y \mapsto \hat{F}(y) := \int_G \rho(g) F(g^{-1} . y)
{\mathrm{d}} \mu_G(g) \in V,
\label{3.6}\end{equation}
where $\mu_G$ denotes the (unique) bi-invariant probability Haar measure
on $G$.
It is easily seen that  $\hat{F} \in C_c^\infty(Y, V)^G$.
It also follows from (\ref{3.6}) that the linear map
\begin{equation}
P \colon C_c^\infty(Y, V) \rightarrow
C_c^\infty(Y, V)^G \subset C_c^\infty(Y, V),
\quad
F \mapsto P F := \hat{F}
\label{3.7}\end{equation}
is a densely defined, symmetric, idempotent, and bounded operator on the
Hilbert space $L^2(Y, V, {\mathrm{d}} \mu_Y)$.
The unique bounded extension of $P$,
$\bar{P} \colon L^2(Y, V, {\mathrm{d}} \mu_Y) \rightarrow
L^2(Y, V, {\mathrm{d}} \mu_Y)^G$,
is just the orthogonal projection onto
${\mathcal{M}} = L^2(Y, V, {\mathrm{d}}\mu_Y)^G$, i.e.,
$\bar{P} = P_{\mathcal{M}}$.
As a consequence, we obtain the relations
$P_{\mathcal{M}} C_c^\infty(Y, V)
= P C_c^\infty(Y, V) = C_c^\infty(Y, V)^G$,
proving the compatibility condition (\ref{2.6})
for $A=\Delta_Y^0$ (\ref{3.3}) and  $S=C_c^\infty(Y,V)^G$.
\rule{5pt}{5pt}
\\[2mm]

\noindent
{\bf Remark 3.1.}
The Hilbert space $L^2(Y, {\mathrm{d}}
\mu_Y)$ naturally carries a continuous unitary representation of $G$.
This is unitarily equivalent to an orthogonal  direct sum,
$L^2(Y, {\mathrm{d}} \mu_Y) \cong \oplus_\rho M_\rho
\otimes V_{\bar \rho}$, where $(\rho, V_\rho)$ runs over a
complete set of pairwise inequivalent
irreducible continuous unitary representations of $G$,
$\bar \rho$ denotes the complex conjugate of the representation
$\rho$, and $M_\rho$ is a `multiplicity space' on which $G$ acts
trivially. Correspondingly, the self-adjoint scalar
Laplace--Beltrami operator $\Delta_Y$, which by definition is the
closure of $\Delta_Y^0$ in (\ref{3.1}), can be decomposed as
$\Delta_Y \cong \oplus_{\rho} \hat\Delta_\rho \otimes
{\mathrm{id}}_{V_{\bar \rho}}$, where $\hat \Delta_{\rho}$ is a
self-adjoint operator on the Hilbert space $M_{\rho}$. It is not
difficult to demonstrate the unitary equivalence
\begin{equation}
(M_\rho, \hat \Delta_\rho)\cong (L^2(Y,V, {\mathrm{d}}\mu_Y)^G, \bar
\Delta_\rho) \quad\hbox{with}\qquad V:= V_\rho,
\label{redsyst}\end{equation} where $\bar \Delta_\rho$ denotes the
closure of $\Delta_\rho$ in (\ref{3.5}). We find it convenient
to use the realization of the \emph{reduced quantum system}
(\ref{redsyst}) furnished by
 $L^2(Y,V, {\mathrm{d}}\mu_Y)^G$
 (see also \cite{TanimuraIwai}).

\section{Explicit description of the reduced systems under polar actions}

We have seen that the \emph{reduced Hamiltonian} entering
the reduced quantum system (\ref{redsyst})
is provided by the essentially self-adjoint
operator $\Delta_\rho$ (\ref{3.5}).
For purposes of interpretation, it
would be desirable to realize the reduced state space
$L^2(Y, V, {\mathrm{d}} \mu_Y)^G$
 as a Hilbert space of
appropriate functions on the reduced configuration space
$Y_{\mathrm{red}} := Y / G$, and the reduced Hamiltonian as a
densely defined differential operator on this space. An apparent
difficulty is that the orbit space $Y / G$ is not a smooth manifold
but a stratified space in general, which among others means that
$Y/G$ is a disjoint union of smooth manifolds of various dimensions.
However, restricting to the \emph{principal orbit type}\footnote{ We
remind that $\check{Y}$ consists of those points $y \in Y$ whose
isotropy subgroups, $G_y$, are the smallest possible for the given
$G$-action; $\check Y$ is open and dense in $Y$. For reviews on
group actions, stratifications and the principal orbit type,
 see e.g.~\cite{Davis,GOV}.},
$\check{Y} \subset Y$, one obtains a \emph{smooth} fiber bundle
$\pi \colon \check{Y} \rightarrow \check{Y} / G$ with fiber
$G / K$ and structure group $N_G(K) / K$, where $K \subset G$ is a
closed subgroup representing the conjugacy class of principal
isotropy subgroups, and $N_G(K)$ stands for the normalizer of $K$ in $G$.
The `big cell'  of the reduced configuration space,
given by $\check{Y}_{\mathrm{red}} := \check{Y} / G$,
is naturally endowed with a
Riemannian metric, $\eta_\mathrm{red}$, making $\pi$ a Riemannian
submersion. From a quantum mechanical point of view, neglecting the
non-principal orbits is harmless, in some sense, since $\check{Y}$ is not
only open and dense in $Y$, but it is also of full
measure.
Indeed,  $\mu_Y(Y \setminus \check{Y}) = 0$ holds, since
$Y \setminus \check{Y}$ is a union of at most countably many
lower dimensional manifolds.

In many interesting applications of Hamiltonian reduction
the group action is \emph{polar}, which means that
it admits \emph{sections} in the sense of Palais and Terng
\cite{PT}.
Recall that a section $\Sigma \subset Y$ is a connected, closed, regularly
embedded smooth submanifold of $Y$ that meets every $G$-orbit and it does
so orthogonally at every intersection point of $\Sigma$ with an
orbit. The induced metric on $\Sigma$ is denoted by $\eta_\Sigma$, and
for the  measure generated by $\eta_\Sigma$ we introduce the
notation $\mu_\Sigma$.
For a section $\Sigma$, denote by $\check{\Sigma}$ a
connected component of the manifold
$\hat{\Sigma} := \check{Y} \cap \Sigma$.
The isotropy subgroups of all elements of $\hat{\Sigma}$ are the same
and for a fixed section we define $K := G_y$ for $y \in \hat{\Sigma}$.
By restricting $\pi \colon \check{Y} \rightarrow \check{Y} / G$
onto $\check{\Sigma}$,
$(\check{Y}_{\mathrm{red}}, \eta_\mathrm{red})$ becomes identified
with $(\check{\Sigma}, \eta_{\check{\Sigma}})$, where $\eta_{\check{\Sigma}}$
is the induced  metric on $\check{\Sigma}$.
We let $\Delta_{\check \Sigma}$ stand for
the Laplace--Beltrami operator of the Riemannian manifold
$(\check \Sigma, \eta_{\check \Sigma})$.
The $G$-equivariant diffeomorphism
\begin{equation}
\check{\Sigma} \times (G / K) \ni (q, g K) \mapsto \phi_g(q) \in \check{Y}
\label{4.2}\end{equation}
provides a trivialization of the fiber bundle
$\pi \colon \check{Y} \rightarrow \check{Y} / G$.
The generalized polar coordinates on $\check Y$
consist of `radial' coordinates on $\check \Sigma$ and  `angular'
 coordinates on $G/K$.

Below we characterize
the reduced system (\ref{redsyst}) in terms of the reduced configuration space
under the simplifying assumption of dealing with polar actions.
We can be brief here as the details, except for the proof
of the essential self-adjointness,
can be found in our recent paper \cite{FP_polar}.

First, let us introduce the linear space
\begin{equation}
\mathrm{Fun}(\check{\Sigma}, V^K)
:= \{ f \, | \, f \in C_c^\infty(\check{\Sigma}, V^K),
f = {\mathcal{F}}|_{\check{\Sigma}} \mbox{ for some }
{\mathcal{F}} \in C_c^\infty(Y, V)^G \},
\label{4.3}\end{equation}
where $V^K$ is spanned by the $K$-invariant vectors in the
representation space $V$. We assume that $\dim(V^K) > 0$.
The restriction of functions appearing in the definition (\ref{4.3})
gives a linear isomorphism
$\mathrm{Fun}(\check{\Sigma}, V^K) \cong C_c^\infty(Y, V)^G \hookrightarrow
L^2(Y, V, {\mathrm{d}} \mu_Y)^G$.
This induces
a scalar product on $\mathrm{Fun}(\check{\Sigma}, V^K)$  making it a
pre-Hilbert space.
The corresponding closure of
$\mathrm{Fun}(\check{\Sigma}, V^K)$  satisfies the Hilbert space
isomorphism
$\mathrm{\overline{Fun}}(\check{\Sigma}, V^K) \cong
L^2(Y, V, {\mathrm{d}} \mu_Y)^G$.

Next,  consider the Lie algebra ${\mathcal{G}} := \mathrm{Lie}(G)$ and its
subalgebra
${\mathcal{K}}:= \mathrm{Lie}(K)$.
Fix a $G$-invariant positive definite
scalar product, ${\mathcal{B}}$, on ${\mathcal{G}}$, and thereby
determine the orthogonal complement ${\mathcal{K}}^\perp$ of
${\mathcal{K}}$ in $\mathcal{G}$.
For any $\xi \in {\mathcal{G}}$
denote by $\xi^{\sharp}$ the associated vector field on $Y$.
Then at each point
$q \in \check{\Sigma}$ the linear map
${\mathcal{K}}^\perp \ni \xi \mapsto \xi^{\sharp}_q \in T_q Y$
is injective, and the \emph{inertia operator}
${\mathcal{J}}(q) \in \mathrm{End}({\mathcal{K}}^\perp)$ can
be defined by the requirement
\begin{equation}
\eta_q(\xi^\sharp_q, \zeta^\sharp_q) =
{\mathcal{B}}(\xi, {\mathcal{J}}(q) \zeta)
\quad
\forall \xi, \zeta \in {\mathcal{K}}^\perp.
\label{4.5}\end{equation}
Note that ${\mathcal{J}}(q)$ is symmetric and positive definite with
respect to
${\mathcal{B}}\vert_{{\mathcal{K}}^\perp\times {\mathcal{K}}^\perp}$.
By choosing dual
bases $\{ T_\alpha \}$, $\{ T^\alpha \}$
$\subset {\mathcal{K}}^\perp$, that is,
${\mathcal{B}}(T^\alpha, T_\beta) = \delta^\alpha_\beta$,
we let
\begin{equation}
b_{\alpha, \beta}(q) := {\mathcal{B}}(T_\alpha, {\mathcal{J}}(q) T_\beta),
\qquad
b^{\alpha, \beta}(q) := {\mathcal{B}}(T^\alpha, {\mathcal{J}}^{-1}(q) T^\beta).
\label{b-def}\end{equation}
We denote the representation of $\mathcal{G}$ corresponding to the
representation $(\rho, V)$ of $G$ as
$\rho' \colon {\mathcal{G}} \rightarrow u(V)$,
where $u(V)$ is the Lie algebra of anti-hermitian operators on $V$.

The $G$-orbit $G . q \subset Y$ through any point $q \in \check{\Sigma}$ is
an embedded submanifold of $Y$ and by its embedding it inherits
a Riemannian metric, $\eta_{G . q}$.
Thus we can define  the (smooth) \emph{density function}
$\delta \colon \check{\Sigma} \rightarrow (0, \infty)$ by
\begin{equation}
\delta(q) := \mbox{volume of the Riemannian manifold } (G . q, \eta_{G . q}),
\label{4.4}\end{equation}
where the volume is understood with respect to
the measure, $\mu_{G\cdot q}$,  belonging to $\eta_{G . q}$.
If $\mu_{G/K}$ denotes the (unique) $G$-invariant probability Haar measure on
$G/K$, then we have the relations
$\mu_{G.q}\cong\delta(q) \mu_{G/K}$ and
\begin{equation}
d\mu_{\check Y} \cong (\delta d\mu_{\check \Sigma}) \times d\mu_{G/K}
\label{extra1}\end{equation}
between the various measures.
Remember that $\check Y \cong \check \Sigma \times (G/K)$
by (\ref{4.2}), and it is also worth pointing out that
$\delta(q) = C \vert \det b_{\alpha\beta}(q)\vert^{\frac{1}{2}}$
with some constant $C>0$.
\\[2mm]

\noindent
\textbf{Proposition 4.1.}
\emph{Consider the reductions of the free particle on $(Y,\eta)$,
with Hamiltonian given by $\Delta_Y^0$ (\ref{3.1}), under a polar $G$-action.
Then, using the above definitions,
the reduced system (\ref{redsyst})
associated with a continuous unitary
irreducible representation $(\rho, V)$ of
$G$ can be identified with the pair
$(L^2(\check{\Sigma}, V^K, {\mathrm{d}} \mu_{\check{\Sigma}}),
\Delta_{\mathrm{red}})$, where
\begin{equation}
\Delta_{\mathrm{red}}
= \Delta_{\check{\Sigma}} - \delta^{-\frac{1}{2}}
\Delta_{\check{\Sigma}}(\delta^{\frac{1}{2}})
+ b^{\alpha, \beta} \rho'(T_\alpha) \rho'(T_\beta)
\label{4.6}\end{equation}
with domain
${\mathcal{D}}(\Delta_{\mathrm{red}})
= \delta^{\frac{1}{2}} \mathrm{Fun}(\check{\Sigma}, V^K)$
is a densely defined, symmetric, essentially self-adjoint operator on
the Hilbert space
$L^2(\check{\Sigma}, V^K, {\mathrm{d}} \mu_{\check{\Sigma}})$.}
\\[2mm]

\noindent
\textbf{Proof.}
We have the Hilbert space identifications
\begin{equation}
L^2(Y, V, {\mathrm{d}} \mu_Y)^G\cong
\overline{\mathrm{Fun}}(\check \Sigma, V^K)
\cong L^2(\check \Sigma, V^K, \delta
{\mathrm{d}}\mu_{\check \Sigma}),
\label{extra2}\end{equation}
where the last equality follows from (\ref{extra1}).
We then consider the isometric isomorphism
$U: L^2(\check \Sigma, V^K, \delta \mathrm{d}\mu_{\check \Sigma})\to
L^2(\check \Sigma, V^K,  \mathrm{d}\mu_{\check \Sigma})$ defined by
the multiplication operator $U$ operating as
$U: f \mapsto \delta^{\frac{1}{2}} f$.
Using $\Delta_\rho$ (\ref{3.5}) and the identifications
(\ref{extra2}),  in \cite{FP_polar}
we have established the explicit formula (\ref{4.6}) for
\begin{equation}
\Delta_{\mathrm{red}}\equiv U \circ \Delta_{\rho} \circ U^{-1}.
\label{extra3}\end{equation}
Hence Proposition 3.1 implies that $\Delta_{\mathrm{red}}$ is
essentially self-adjoint on the domain obtained by transferring the
domain of $\Delta_{\rho}$ into
$L^2(\check{\Sigma}, V^K, {\mathrm{d}} \mu_{\check{\Sigma}})$ by means
of (\ref{extra2}) and the map $U$.
The  so-obtained domain is just
$\delta^{\frac{1}{2}} \mathrm{Fun}(\check{\Sigma}, V^K)$.
\rule{5pt}{5pt}
\\[2mm]

\noindent
{\bf Remark 4.1.}
The first term of formula (\ref{4.6})
corresponds to the kinetic energy of a particle moving
on $(\check Y_{\mathrm{red}}, \eta_{\mathrm{red}})\cong
(\check \Sigma, \eta_{\check \Sigma})$  and the rest
represents potential energy if $\mathrm{dim}(V^K)=1$
(which happens very rarely \cite{EFK,Obl}).
The second term of (\ref{4.6}) is of course always potential
energy (and in certain cases just a constant).
If $\mathrm{dim}(V^K)>1$,
then one says that the reduced system contains internal `spin'
degrees of freedom and
the last term of (\ref{4.6}) encodes spin-dependent potential energy.
For comparison with the result of classical Hamiltonian
reduction, see \cite{FP_polar} and references therein.

\section{Conclusion}

In this work we focused our attention on
the inheritance of the  (essential) self-adjointness
property in the process of quantum Hamiltonian reduction.
By using the auxiliary result of Section 2, we proved in Section 3
the essential self-adjointness of the reduced Laplace--Beltrami
operators (\ref{3.5})  on complete Riemannian manifolds
equipped with smooth isometric actions of compact Lie groups.
Lemma 2.1  can
be applied in other examples as well to prove essential
self-adjointness for reduced Hamiltonians.

Assuming that the symmetry group acts in a polar manner,
in Section 4 we recalled
the explicit realization (\ref{4.6})
of the reduced Laplace--Beltrami operators
in terms of the reduced
configuration space, and complemented the results of \cite{FP_polar}
by verifying the essential self-adjointness of these operators.
It is worth noting that if the underlying  Riemannian manifold is
itself a Lie group, or a symmetric space,
and the `sections' can be realized as flat Abelian subgroups, then the
Hamiltonian reductions of the Laplace--Beltrami operator typically
yield spin Calogero--Sutherland type models
(see e.g. \cite{OPII,EFK,Obl,FP1,FP3}).
Many known and new (non-elliptic) spin
Calogero--Sutherland type models arise in
this framework and certain spinless cases
also appear among the reduced systems.
We shall further elaborate the structure of
these models in the future.
\\[2mm]

\noindent \textbf{Acknowledgments.} The work of L.F. was supported
by the Hungarian Scientific Research Fund (OTKA grant
 T049495)  and by the EU network `ENIGMA'
(contract number MRTN-CT-2004-5652). B.G.P. is grateful to J. Harnad
for hospitality in Montr\'eal.

\end{document}